\documentclass[pre,amsmath,amssymb,twocolumn,superscriptaddress]{revtex4-2}
\usepackage{amsmath,amssymb}
\usepackage{mathtools}
\usepackage[usenames]{color}
\usepackage{amssymb}
\usepackage{dsfont}
\usepackage{grffile}
\usepackage{textcomp}
\usepackage{xspace}
\usepackage{hyperref}
\usepackage{graphicx} 
\usepackage{outlines}
\usepackage{natbib}
\hypersetup{
        citecolor  = blue,
        colorlinks = true,
    }

 \usepackage{dsfont}
\usepackage{graphicx}
 \usepackage{amsmath, amssymb}
 \usepackage{bbold}
 \usepackage{braket}
\usepackage{hyperref}
 \usepackage{physics}

\newcommand{\la}{\langle}
\newcommand{\ra}{\rangle}

\newcommand\Mycomb[2]{\prescript{#1\mkern-0.5mu}{}C_{#2}}

\newcommand{\smt}{Science, Mathematics and Technology Cluster, Singapore University of Technology and Design, 8 Somapah Road, 487372 Singapore}
\newcommand{\esd}{ESD Pillar, Singapore University of Technology and Design, 8 Somapah Road, 487372 Singapore}
\newcommand{\epd}{EPD Pillar, Singapore University of Technology and Design, 8 Somapah Road, 487372 Singapore} 
\newcommand{\cqt}{Centre for Quantum Technologies, National University of Singapore 117543, Singapore}

\newcommand{\id}{\mathcal{I}} 
\newcommand{\avH}{\langle H \rangle} 
\newcommand{\im}{{\rm i}}
\newcommand{\Loss}{\mathcal{L}}

\begin{document} 


\title{Entangling capabilities and unitary quantum games}


\author{Rebecca Erbanni}
\affiliation{\smt} 
\author{Antonios Varvitsiotis} 
\affiliation{\esd} 
\author{Dario Poletti}
\affiliation{\smt} 
\affiliation{\epd} 
\affiliation{\cqt} 
\affiliation{MajuLab, International Joint Research Unit UMI 3654,
CNRS-UCA-SU-NUS-NTU, Singapore}

\begin{abstract}
    We consider a class of games 
    between two competing players that take turns acting on the same many-body quantum register. Each player can perform unitary operations on the register, and after each one of them acts on the register the energy is measured. Player A aims to maximize the energy while player B to minimize it. 
    This class of zero-sum games has a clear second mover advantage if both players can entangle the same portion of the register. We show, however, that if the first player can entangle a larger number of qubits than the second player (which we refer to as having quantum advantage), then the second mover advantage can be significantly reduced. 
    We study the game for different types of quantum advantage of player A versus player B and for different sizes of the register, in particular, scenarios in which absolutely maximally entangled states cannot be achieved. In this case, we also study the effectiveness of using random unitaries. Last, we consider mixed initial preparations of the register, in which case the player with a quantum advantage can rely on strategies stemming from the theory of ergotropy of quantum batteries. 
\end{abstract}

\maketitle

\section{Introduction} 

Game theory is a very impactful branch of mathematics with applications in economics, political sciences, computer science etc. Within this theory one studies the interaction between, and choices of, different players which may be involved in a cooperative or competitive game \cite{Tadelis2013}. A natural development of game theory is that of quantum game theory, in which the rules of the game, and what players can do, are prescripted within the limit of quantum physics and not just classical physics \cite{Meyer1999, EisertLewenstein1999, MarinattoWeber2000, EisertWilkens2000, BenjaminHayden2001, BenjaminHayden2001b, Johnson2001, KayBenjamin2001, DuHan2002, FlitneyAbbott2002, MarriottWatrous2005, GutoskiWatrous2007,  Frackiewicz2011, Zhang2012, GuoKoehler2008, Jain2009, Jain2014}. 
For instance, it is possible to design a high-payoff coherent quantum equilibrium in a prisoner dilemma's type of game \cite{BenjaminHayden2001b}, which would be significantly affected by the presence of noise \cite{Johnson2001}. Experiments on quantum games in which one player can perform quantum operations have also been demonstrated \cite{ZuDuan2012}. 

Here we consider a competitive zero-sum game in which players A and B are acting on a single many-body quantum system and each is trying to obtain a different objective. 
Importantly, because of the physics of the game considered we deal with a turn-based, or sequential, game, where players take turn to make their moves. This brings this study outside of the typical realm of von Neumann's theorem \cite{Neumann1928}, in which the action of the players are simultaneous. 
Instead, here, similarly to Stackelberg games developed to study the evolution of economic systems in which a firm moves first and it is followed by its competitors \cite{stackelberg}, the outcome of the game can be significantly affected by the order in which the players act. 
For instance, the second player may know exactly what the first player has done, hence the first player needs to strategize knowing the second player can do the best counter-action to his/her first action. A simplified scenario is that of duo-poly, where only two players are in the game \cite{osborne1974duopoly,eliashberg1991competitive,huck2001stackelberg}.         
Depending on the details of the game, two possible scenarios can emerge: a {\it first mover advantage} \cite{annen2019first}, which stems from the fact that the first player significantly affects the system in a way that limits what the second player can do, and a {\it second mover advantage} \cite{amir2006second} in which the second player can readily counteract the first player's move and change the situation in his/her favor. 
The Stackelberg duopoly scenario has been studied in quantum systems too, where it was shown the ability to entangle states, the presence of noise and memory, can affect the positional advantage \cite{LoKiang2003, LoKiang2005, WangXia2007,  ZhuXia2008, Khan2010, FracPykac2018, AlonsoMartin2020, ShiXu2021, ShiChen2021}.       
\begin{figure}[h!]
    \centering
    \includegraphics[width=0.4\textwidth]{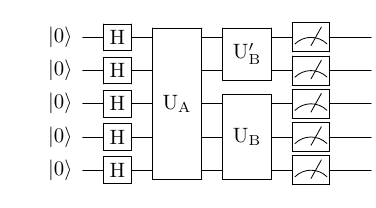}
    \caption{Circuit representation of the sequential two-player one-step game studied in this work. The register is initialized in a product state as defined in Eq (\ref{eq:prod_state}), that is acted upon by both players with non-commuting unitaries in a sequential manner. Player A tries to maximize the energy while B to minimize it. Then the measurement outcome of the energy of Hamiltonian (\ref{eq:ham}) assigns the win. 
    }
    \label{fig:plot0}
\end{figure}

We consider a game in which, 
for players with equal abilities, the second mover has a significant advantage. However we show that if one player has the ability to entangle and operate on larger portions of the many-body quantum system, he/she would have a significant advantage and may not lose even if he/she is moving first, thus completely erasing the second mover advantage. We refer to this ability to entangle and act on larger portions of the system as ``quantum advantage''. A depiction of this scenario is represented in Fig.~\ref{fig:plot0} where player A can entangle 5 qubits, while player B can entangle at most 3. In this manuscript, we show in which scenario the first player can play what we refer to as {\it perfect defence}, where the second player can only obtain a draw despite moving second. This is possible when the first player generates maximally entangled states. We also show how close the first player can be to the perfect defence when this cannot be implemented. Importantly, this advantage would not be possible in a directly corresponding classical game. 
Last, inspired by recent results on the theory of quantum batteries and ergotropy \cite{Allahverdyan2004, AlickiFannes2013}, we show further strategies that the player with a quantum advantage can implement in the case of an initial preparation of the quantum system/register in a mixed state.  

The paper in structured in the following manner: in Sec.~\ref{sec:game_setup} we introduce the game in detail, and in Sec.~\ref{sec:winning_strategy} we show how the player with a quantum advantage can build a winning strategy, in different scenarios for different system sizes, dimension of local Hilbert space, and different abilities to operate on portions (or totality) of the system. In Sec.~\ref{sec:classical} we make a comparison with a classical corresponding case, and then we consider an initial preparation of the system in a mixed state in Sec.~\ref{sec:mixed_states}. In Sec.~\ref{sec:conclusions} we draw our conclusions.

\section{Setup of the unitary quantum game}\label{sec:game_setup} 

The scenario we consider is that in which two players, Alice (A) and Bob (B), can act on a single many-body quantum state. For most of this work we consider this state to be pure, and initialized as a product state of $N$ qubits. 
The purpose of player A (B) is to maximize (minimize) the expectation value of the energy which we define as $\langle H \rangle = \bra{\psi}H\ket{\psi}$ where 
\begin{align}
    H = \sum_{n=1}^N \sigma^z_n,  
\label{eq:ham}
\end{align}
which is simply the sum of local Pauli $\sigma^z_n$ operators on each qubit $n$. This choice of $H$ to evaluate the payoffs makes this a zero-sum game. 

Each player can only act once with a single unitary \footnote{Excluding the possibility that a player applies multiple layers of unitaries.} on the system, and the players take turns to do their move. Both players know perfectly what the state of the system is and, if they are the second player to act on $\ket{\psi}$, they also know what unitary the other player has used to act on the system. 

In this scenario if player A and B can apply the same type of unitaries, then it is obvious that the player who acts second will have a clear advantage. 
Suppose for instance that player A acts first and player B second. For whichever unitary $U_A$ player A executes, player B can operate with $U_B U_A^\dagger$ thus first completely removing the effects of A's unitary, and then acting with the unitary $U_B$ which is suited to minimize $\langle H \rangle$. 
For this reason, in games of this type one is inclined to speak of {\it second mover advantage}. For clarity, in the following we will use the notation $\avH_{AB}$ when player A acts first and player B second, and $\avH_{BA}$ in the opposite case.  

In this work we consider the scenario in which player A has a quantum advantage. 
More precisely, player A can perform unitaries on a number of qubits larger than B, for instance up to $N_A=N$ qubits, while player B can only act on, at most, $N_B<N$ qubits.  
Hence, the operations performed by player $A$ can entangle more qubits than those of $B$. 

As an initial state we consider a product pure state such that $\langle H\rangle = 0$, for instance 
\begin{equation}
    | \psi_0 \rangle = \otimes_{l=1}^N \frac{ \left(|0\rangle_l + |1\rangle_l \right)} {\sqrt{2}}.   
\label{eq:prod_state}
\end{equation}
It should be noted, though, that the details of the initial pure state are not important. 

\section{Comparison with the corresponding classical game} \label{sec:classical} 
To better gain an insight on the role of quantum effects, we first consider a corresponding classical game which, as we will see soon, is uneventful. 
In this corresponding scenario we consider that both the preparation of the initial state and the operation acting on the state are deterministic. Hence, for this classical game the register will only contain 0s or 1s, and no superpositions of them. Each player can only flip the spins or keep them as they are. On each realization of the problem, the second player, knowing the initial preparation of the register, and the actions implemented by the first player, can readily figure out which bits to flip in order to obtain the preferred configuration (in fact, the second player can even do measurements and choose its move accordingly). It results that the second player has a clear second mover advantage which cannot be overcome by the first player.     
Hence, the game only becomes interesting when superposition and entanglement between different parts become possible, i.e. in the quantum regime. 

Another major difference between the classical and quantum game is that in the classical case the energy measured from the final state of the register is deterministic while for the quantum case this is intrinsically probabilistic; in the classical case, repeating the game numerous times will not change the fact that the second player will consistently obtain the best gain possible, while for the quantum game, although the protocols are deterministic, for a given realization of the problem there could be different measured values, and the results based on $\avH_{AB}$ should only be understood on average.      

\section{Player's A principle for a winning strategy in the quantum game}\label{sec:winning_strategy} 
Here we are going to describe the main principle following which player A, who has a quantum advantage, can have a winning strategy. Since player B wants to minimize the energy, player A, when acting first, tries to maximize the minimum energy that B can reach. In other words, player A is doing a {\it max-min} computation typical of games between two players.    
We exemplify it at first using the case of $N=N_A=2$ and $N_B=1$. 
In this case player A can immediately neutralize player B by transforming $\ket{\psi_0}$ into, for instance, the maximally entangled state 
\begin{equation}
    \ket{\psi_{A}}= \frac{\ket{00}+\ket{11}}{\sqrt{2}}.  \label{eq:max_entangled}  
\end{equation} 
Then player B, who is only able to apply unitaries on a single qubit, is actually trying to minimize $\langle H\rangle_{AB} $ 
\begin{align}
    \avH_{AB} = & \tr\left( H U_B\ket{\psi_{A}}\bra{\psi_{A}}U_B^\dagger\right) \nonumber \\ 
    & = \sum_l \tr\left( \sigma^z_l U_B\ket{\psi_{A}}\bra{\psi_{A}} U_B^\dagger\right) \nonumber \\ 
    & = \sum_l \tr\left( \sigma^z_l \frac{\id_l}{2}\right) = 0 \label{eq:2qubitenergy}
\end{align}
where $\id_l$ is the identity matrix which results from the partial trace of $\ket{\psi_{A}}$.  
It is thus impossible for B to change the expectation value of the energy for whichever unitary $U_B$ that B may choose to apply.  
If A instead plays as the second player, then she can always undo what B has done and thus maximize $\langle H \rangle_{BA}$ with a unitary $U_A U_B^\dagger$. 
Player A has, then, a perfect defensive strategy when it plays first which does not allow player B to win, and a winning strategy when acting second. 
Any other strategy would allow B to produce an energy lower than $0$, which would be less ideal for player A. 

The above strategy can be generalized to more qubits, although with care. For now, we still consider $N=N_A>N_B\ge N/2$. After player A has acted with a unitary $U_A$, one can write 
\begin{equation}
\ket{\psi_A}=\sum_{i=1}^{\min\{d_D,d_B\}} \sqrt{\lambda_i}\ket{\alpha_i}_D\ket{\beta_i}_B
\end{equation}
where $d_D = 2^{N_A-N_B}$ and $d_B = 2^{N_B}$.  

The reduced density matrix over the $N_B$ qubits then is  
\begin{align}
    \rho_B&=\sum_i^{\min\{d_D,d_B\}} \lambda_i\ket{\beta_i}_B\bra{\beta_i}_B \nonumber \\ 
    &=\begin{bmatrix} \lambda_1 & 0 & .. & 0& .. & ..&..&  0 \\
       0 & \lambda_2 & .. & 0  & .. & ..&..&  0 \\
        .. & .. & ..  &.. & .. & .. & ..& ..\\
         0 & 0 & .. & \lambda_{\min\{d_D,d_B\}} & .. & ..& ..& 0 \\
         0 & 0 & .. & 0 & 0 & .. & ..&  0 \\
         0 & 0 & .. & 0 &.. & 0 &..&  0 \\
         .. & .. & ..  &.. & .. & .. & ..& ..\\
         0 & 0 & .. & 0 &.. & .. &..&  0 \\
     \end{bmatrix} \nonumber 
\end{align}
where the number of non-zero $\lambda_i$ is the Schmidt rank. 

As player B acts with a unitary on $N_B$ qubits, he cannot change the values of the eigenvalues $\lambda_i$. This limits player B's ability to lower the energy. In fact, the best unitary operation that player B can do is to turn that mixed state $\rho_B$ into the passive state $\rho^P_B$, i.e. 
\begin{align}
    \rho^P_B = \sum_m \tilde{\lambda}_m |E_m\rangle, \label{eq:passive_B}    
\end{align}
where $|E_m\rangle$ are the energy eigenstates of $H$ in an increasing value of the energy, while the $\tilde{\lambda}_m$ are a decreasing ordering of the eigenvalues $\lambda_i$ \cite{Allahverdyan2004}. We remind the reader that a state $\rho$ is deemed passive for an Hamiltonian $H$ iff $\tr(H U\rho U^\dagger)\ge\tr(H \rho)$ for any unitary $U$.     

It is thus clear that player A's best strategy is to limit as much the effects of any action that player B can do, and this is achieved by trying to set the eigenvalues $\lambda_i$ to be equal or, in other words, increase the entropy of the reduced density matrix $\rho_B$. In particular, as we have seen for the two qubits example, if player A does an operation such that $\rho_B$ is proportional to the identity matrix, any operation $U_B$ that player B will do is ineffective. 
In the following sections we can see the scaling of the gains that player A can have when playing against player B as a function of the system size and of the relative quantum advantage of player A versus player B. 
Before continuing, though, we note that if player B cannot even entangle $N/2$ qubits, then his actions cannot be better.

\subsection{Player A with minimal advantage over B} \label{sec:minimal advantage}    
It is natural to think that if player A and player B can address a very similar number of qubits, than the advantage of player A versus player B will be limited. In this section we start to explore this aspect. 
The smallest advantage that player A can have over player B is to be able to entangle just one more qubit. In this case, hence, $N_A=N$ and $N_B=N-1$. 
Since the case in which player A acts second is trivial, we only consider the case in which player A acts first. In this case player B can choose any $N_B$ qubits to operate a large unitary $U_B$, and then there will be a left-over qubit over which player B will apply a single qubit unitary.  
Considering the state $\ket{\psi_A}$ prepared by player A, from the Schmidt decomposition it is easy to see that there are at most two non-zero values of $\lambda_i$ and thus the best strategy for player A is to give to player B a mixed state in which these two eigenvalues are identical, i.e.  
\begin{equation}
    \rho_B=\sum_i^{2} \frac 1 2 \ket{\beta_i}_B\bra{\beta_i}_B= \frac 1 2 \begin{bmatrix} 1 & 0 & 0 & 0& .. & 0 \\
       0 & 1 & 0 & 0  & ..& 0 \\
        0 &0 & 0 & 0  & .. &  0 \\
        .. & .. & ..  &.. & .. & ..\\
         0 & 0 & 0 & 0 & .. &  0 \\
         0 & 0 & 0 & 0 & .. &  0 \\
         0 & 0  &0 & 0 &..&  0 \\
     \end{bmatrix}. \label{eq:rho2}
\end{equation}
Here we have implicitly assumed that player A can actually generate this state $\ket{\psi_A}$ for any choice of the $N_B$ qubits from player B. 
But, as we will see in more detail later, this is generally not the case. 

At this point, player $B$ can only act on the two reduced density matrices, one over $N_B$ qubits and the other over a single qubit. From the second one, player $B$ cannot act meaningfully with any unitary, as it is proportional to the identity matrix. For the first one, instead, player B is dealing with a matrix with rank $2$ and two equal eigenvalues with value $1/2$. Since with unitary operations player B cannot change the eigenvalues of the reduced density matrix, the minimum energy he can reach is obtained considering the two lowest eigenvalues of energy available, i.e. $-N_B$ and $-N_B + 2$ respectively for all the $N_B$ spins pointing down, or all expect for one. This results in a value of $\langle H \rangle$ per qubit which is 
\begin{equation}
    \frac{\langle H \rangle_{AB}} {N} = - \frac{1}{N}\left[\frac{N_B}{2} + \frac{1}{2}(N_B-2)\right]= -1 + \frac{2}{N}.  
\label{eq:en1}
\end{equation}
It thus results that the larger $N$ is, the least player A will be able to affect player B and the more player B will be able to minimize the energy of the Hamiltonian. We can see that Eq.~(\ref{eq:en1}) is consistent with the result obtained for $N=2$, for which $\langle H \rangle_{AB}=0$, see Eq.~(\ref{eq:2qubitenergy}). This, of course, provided that player A can actually set B into the scenario described in Eq.~(\ref{eq:rho2}), but as we will see later, this may not always be possible.

\subsection{General number of qubits and absolutely maximally entangled states}\label{sec:general}          If the difference in system sizes is $N-N_B=M$, then the density matrix $\rho_B$ will have a number of identical and nonzero eigenvalues given by the minimum between $2^{N_B}$ and $2^M$, whose sum is one. If $M \ge N_B$ then player B could receive a completely mixed state proportional to the identity matrix. 
\begin{figure}[ht!]
\includegraphics[width=0.93\columnwidth]{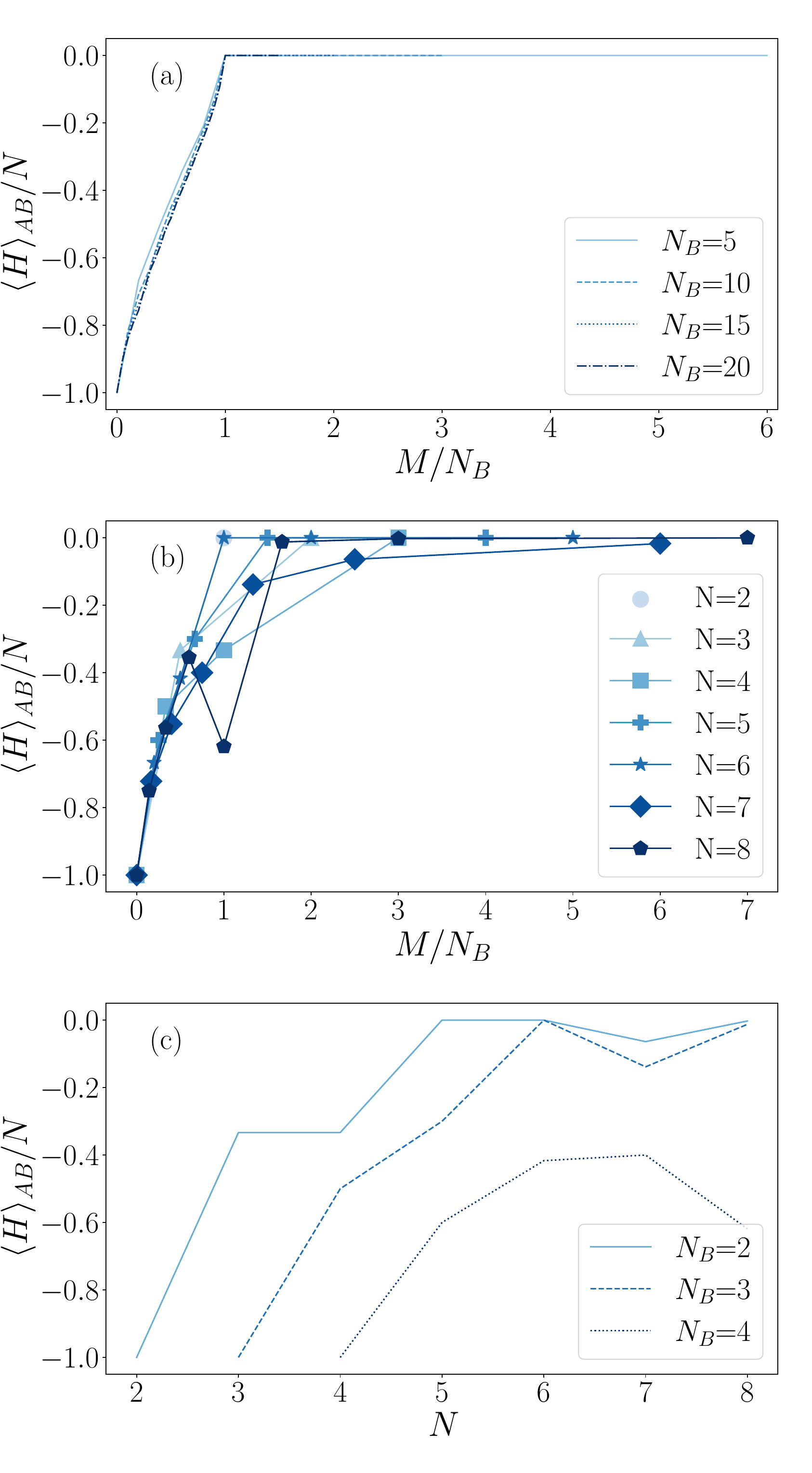}
    \caption{(a) Plot of $\avH_{AB}/N$ from Eq.~(\ref{eq:formulaH2}) versus $M/N_B$ for different values of $N_B$ assuming that player A can always produce AMEs. (b) Similar to panel (a) except that here we consider the states with maximal entanglement that $A$ can produce and we plot the minimum value of $\avH_{AB}$ that player B can achieve considering all different partitions of the qubits. Each curve corresponds to a different value of $N$ and we compute the energies for $N_B \in (1,N)$. (c) Plot of player B's achievable minimum value of $\avH_{AB}/N$ as a function of the total number of qubits $N$ when $N=N_A$, for different sizes of system B, $N_B$. Each line starts from $N=N_B$ and goes until $N=8$.}
\label{fig:plot2_3} 
\end{figure} 
For $M < N_B$ it is possible to compute the average energy per qubit for player B when he is the second to act, which is given by associating the bottom $2^M$ energy levels (which can be highly degenerate) with an occupation $1/2^M$ and results in: 
\begin{equation}
    \frac{\langle H\rangle_{{AB}}}{N}=-\frac{1}{N}\sum_{i=1}^{2^{N_B}}\left[ \sum_{k=0}^{N_B}  (N_B-2k) v_i u_i \right]
\label{eq:formulaH2}
\end{equation}
with 
\begin{equation}
    v_i=\left[\sum_{l=0}^{k-1}\Mycomb{N_B}{l}< i\leq \sum_{l=0}^{k}\Mycomb{N_B}{l}\right]_I 
\label{eq:vi}
\end{equation}
\begin{equation}
    u_i= \frac{1}{2^{M}}[i\leq 2^M]_I
\label{eq:ui}
\end{equation}
where $[\cdot]_I$ denotes the Iverson bracket that equals 1 when the argument is true and 0 when it is false, and $\Mycomb{n}{m} = \frac{n!}{(n-m)!m!}$ \footnote{Note that if $m=0$ then there are no terms in the sum, and hence it gives 0.} is the number of different configurations of $n$ spins with $m$ spins down.

The minimum energy that player $B$ can produce is thus represented in 
Fig.~\ref{fig:plot2_3}(a). Each line corresponds to a different value of $N_B=5,\;10,\;15$ and $20$. As $N_B$ increases the curve converges towards a smooth function which has a discontinuity in the derivative as it approaches $M=N_B$. 
%

For this to happen, however, player A would need to be able to generate an absolutely maximally entangled state (AME) for any number of qubits $N$, i.e. the reduced density matrix of any subsystem of $k$ qubit is the maximally mixed state \cite{Enriquez_2016, Helwig_2012,Facchi_2008,Goyeneche_2015}, where $k\leq\lfloor N/2\rfloor$ and $\lfloor \;\cdot\;\rfloor$ refers to the floor function \footnote{See also applications of AME in quantum error correction \cite{Scott_2004,huffman2010fundamentals,raissi2018optimal}}. 
In practice, we will focus on the equality $k=\lfloor n/2\rfloor$, since if we can obtain the maximally mixed state on $\lfloor n/2\rfloor$ qubits, then any subsequent marginal density matrix will also be maximally mixed. 
AME states for 2 and 3 qubits are simply the classes of Bell and GHZ states, while for 4 qubits, an AME state does not exist \cite{Higuchi_2000,Gour_2010}. For 5 and 6 qubits, AME states have been found in \cite{Facchi_2008} by minimizing the average purity of every bipartition i.e. $\frac{1}{N}Tr_B(\rho^{2})$, while for $N=7$ \cite{Huber_2017} and $N\geq8$ \cite{Scott_2004}, the frustration between different subsystems makes it impossible to reach the maximum entropy over all marginal density matrices  \cite{Facchi_2010b}. 
Hence, for qubits one can only produce AMEs for $N=2,3,5,6$. We note, however, that one can still find AMEs states for any $N$ by increasing the number of levels of each sub-system, e.g. using qudits, where $d$ indicates a number of levels larger than 2 \footnote{An updated table of all known AME states is available online \cite{table_AMEs}}. 

Still considering only qubit systems, we can numerically search for the AME states or closest states to them, in order to evaluate the minimum value of $\avH_{AB}$. Our results confirm previous numerical searches \cite{Facchi_2008,Borras_2007,Zha_2012}. 
Given a parametrized ansatz for the wave function, we minimize the mean von Neumann entropy over any subsystem K of $k$ qubits, where the von Neumann entropy and the reduced density matrix are defined as 
\begin{equation}  
    S(\rho_{K})=-\tr\left[\rho_{K} \log_d(\rho_{K})\right], 
\end{equation}
where $d$ is the size of each system, here 2 because we are dealing with qubits, 
and 
\begin{equation}
    \rho_{K}=\tr_{\overline{K}}(\ket{\psi}\bra{\psi}). 
\end{equation}
Here $(K,\overline{K})$ defines a bipartition of the $N$ qubits, respectively with $k$ and $N-k$ qubits.\\
We then choose the loss function to be minimized as the mean of the von Neumann entropy over the different subsystems  
\begin{equation}
    \Loss=-\frac{1}{L}\sum_{i=1}^{L} S(\rho_K)
\end{equation}
with $L=\frac{N!}{(N-k)!k!}$ if $N$ is odd and $L=\frac{N!}{2(N-k)!k!}$ if it is even since, in this case, the entropies over K and $\overline{K}$ are equal. 

For the wave function, we use a symmetrical ansatz for 4 qubits 
%
\begin{align}
    \ket{\psi}&=a_0 (e^{i\theta_0}\ket{0000} +e^{i\theta_1}\ket{ 1111}) \nonumber \\ 
    & + a_1 (e^{i\theta_2}\ket{0001} + e^{i\theta_3}\ket{0010} + e^{i\theta_4}\ket{0100}+ e^{i\theta_5}\ket{1000} \nonumber \\ 
    & + e^{i\theta_6}\ket{0111} + e^{i\theta_7}\ket{1011} + e^{i\theta_8}\ket{1101} + e^{i\theta_9}\ket{1110}) \nonumber \\ 
    &+ a_2 (e^{i\theta_{10}}\ket{0011} + e^{i\theta_{11}}\ket{0110} + e^{i\theta_{12}}\ket{1100} \nonumber \\        
    &+  e^{i\theta_{13}}\ket{0101} + e^{i\theta_{14}}\ket{1010} + e^{i\theta_{15}}\ket{1001} )
\label{eq:4qb_ansatz}
\end{align}
where $a_i=0,1$ and perform a minimization routine for each of the 7 combinations of $a_0,a_1,a_2$. With this ansatz, we are indeed able to retrieve the result of \cite{Higuchi_2000} which was proven to be a local maximum of the von Neumann entropy-based loss \cite{Brierley_2007}, on top of finding the AMEs for $N=2,3,5,6$. 
In Fig.~\ref{fig:plot2_3}(b) we show $\avH_{AB}/N$ versus $M/N_B$ for a number of qubits up to $N=8$ for player A, generating the state with maximum average entanglement between different partitions, which gives a more realistic realization of what was shown in Fig.~\ref{fig:plot2_3}(a). In Fig.~\ref{fig:plot2_3}(c), for clearer insight, we show the data in a different form with $\avH_{AB}/N$ versus $N$ for different values of $N_B$. Each line starts from $N=N_B$ and goes until $N=8$. For $N$ such that there is an AME, then $N_B$ needs to be larger than $N/2$ for $\avH_{AB}/N$ to be lesser than $0$. For $N\ne 2,3,5,6$, it is possible for different values of $N_B$ to have negative $\avH_{AB}/N$, although it may not reach $-1$.

\begin{figure}[h!]
  \includegraphics[width=0.8\columnwidth]{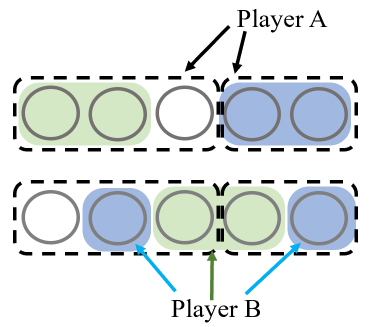}
  \caption{(a) Player A entangles qubits 1 to 3 and 4 to 5, while player B entangles qubits 1-2 and 4-5. (b) Player A entangles qubits 1 to 3 and 4 to 5, while player B entangles qubits 3-4 and 2-5. Common parameters: $N=5, N_A=3, N_B=2$.}
  \label{fig:case1}
\end{figure} 

\subsection{Player A cannot act on the whole system}\label{sec:notwhole}
Now we consider the setting where player A can entangle only a subset of the N qubits, while still having an advantage compared to player B, i.e.  $N_B<N_A<N$. 
To gain an insight, we first consider a system of $N=5$ qubits, $N_A=3$ and $N_B=2$. We take the case that player A decides to entangle qubits 1,2,3 together and qubits 4,5 together. Then player B can choose to act on different pairs of states. For player B the best course of action is to act separately on qubits 1, 2 together and 4, 5 together and then on qubit 3 as pictured in Fig.~\ref{fig:case1}. On qubits 4 and 5, player B can perfectly reverse player A's action, thus getting the minimum energy of -2 from this block. 
Howewer, player $B$ cannot completely alter the state of qubits 1 to 3 because player A has turned them into an AME. 
More specifically, qubit 3 will be in the completely mixed state, while qubits 1 and 2, for player B, are in the following diagonal state:
\begin{equation}
    \rho_{1,2}=\frac{1}{2}\begin{pmatrix}
    1       &  & &  \\
    & 1 & &   \\
    & & 0&   \\
   & & &0   \\
\end{pmatrix}. 
\end{equation}
Hence the minimum value of $\avH_{AB}$ that player B can obtain is $\avH_{AB}/N=-3/5$. This is a clear improvement for player B compared to the case in which $N_A=3$ and $N_B=2$ as analyzed before for $N_A=N_B+1$, but with $N=N_A$, which would give $\avH_{AB}/N=-1/3$, see App.~\ref{app:example} for more details.

The value $\avH_{AB}/N=-3/5$ is also the best possible outcome for player B when $N=5$, $N_A=3$ and $N_B=2$. In fact, player B could choose other pairs of qubit, e.g. qubits 3,4 and qubits 2, 5, as depicted in Fig.~\ref{fig:case1}(b), or equivalent permutations, and their respective reduced density matrices would be 
\begin{equation}
    \rho_{2,5}=\rho_{3,4}=\frac{1}{2}I_2\otimes \frac{1}{2}I_2=\frac{1}{4}I_4
\end{equation}
which give $\avH_{AB}=0$. 
From this simple example we learn that, if $N_A<N$, player $B$ has further chances to reduce the gap between $\avH_{AB}$ and $\avH_{BA}$ despite $N_A>N_B$. 

\subsection{Using random unitaries}\label{sec:random_unitaries}     
For larger registers, a strategy that player A could perform is to generate a highly entangled state by applying random unitaries drawn from the Haar distribution \cite{Goyeneche_2015}, i.e. the uniformly random distribution over the unitary group $\mathit{U}(N)$ of matrices of size $N\times N$. 
The average entropy will then grow linearly with the number of qubits \cite{Goyeneche_2015}.

We consider the case of four qubits, which we, at first, set in the zero state, and then we sample by applying a random unitary from $\mathit{U}(N)$ and check the mean entropy over the three possible $2-$qubit subsystems. 
Considering 1000 samples, we obtain a maximum for the expectation value of the subsystem entropy of $\approx$1.64, and overall a mean value $\approx$1.33 with standard deviation $\sigma\approx 0.12$. Note that the maximum entropy averaged over the two 2-site subsystems for a $4-$qubit state is $\approx 1.79$ \cite{Higuchi_2000}. 
While it is not possile to reach an AME, and it is also difficult to reach the largest possible value, by sampling random unitaries one can produce 2-site reduced density matrix with an average of 1.33. This implies that player B, acting second, cannot completely reverse what player A has done and thus cannot reach the minimum possible value of $\langle H \rangle_{AB}$. If player A and player B play the game many times and equally alternate who plays first, on average A will be able to win thanks to her quantum advantage.

\section{Mixed states and maximization of ergotropy}\label{sec:mixed_states} 

Here we want to show that if player A has a quantum advantage, she can further improve on the extracted energy for mixed systems with a local Hilbert space larger than two. To explain this, it could be helpful to do a little detour. In the study of quantum batteries, it has been shown in \cite{AlickiFannes2013} that minimizing the energy of a collection of identical systems can lead to a single-system energy which is lower than what one would obtain if he/she was to minimize the energy of each single system individually. In particular, this is possible when the energy distribution in each single system is not thermal. To be more specific, if we consider a system given by 
\begin{align}
    \rho = \prod_i \rho_i \label{eq:mixed_initial}
\end{align}
where $\rho_i$ is a diagonal matrix, and now we apply the same single-site unitary on each density matrix to maximize its energy, we can obtain an expectation value of the energy per particle which we refer to as $\avH_s/N$. If instead we can apply a unitary on many sites, even if we just measure single-site, local, energies, it is possible to obtain an expectation value of the energy $\avH_m$ such that 
\begin{align}
\frac{\avH_m}{N}>\frac{\avH_s}{N}. \label{eq:ergotropy_scaling}      
\end{align}
For this to occur, the $\rho_i$ should not be thermal. Hence we cannot consider two-level systems, as their diagonal matrices can always be considered as in a thermal state. We also want to start from an initial state for which $\avH=0$, i.e. unbiased between player A and player B, and which can satisfy Eq.~(\ref{eq:ergotropy_scaling}). The minimum size of the local Hilbert space which allows this to happen is 5. We thus start from a product of local density matrices as in Eq.~(\ref{eq:mixed_initial}) such that the local ones can be written as 
\begin{align}
\rho_i=\begin{pmatrix}
    p_2       &  & & & \\
    & p_1 & & &  \\
    & & p_0 & &  \\
   & & & p_1 &   \\
   & & & &  p_2    \\
\end{pmatrix}. 
\end{align}
If the Hamiltonian of the system is 
\begin{align}
H=\begin{pmatrix}
    E_2       &  & & & \\
    & E_1 & & &  \\
    & & 0 & &  \\
   & & & -E_1 &   \\
   & & & &  -E_2   \\
\end{pmatrix}
\end{align}
with $E_2>E_1> 0$, then the maximum energy that one can obtain from applying a unitary on a single site $\avH_1$ is, considering (without loss of generality) $0<p_2<p_1<p_0$, 
\begin{align}
    \avH_1 = p_0 E_2 + p_1 E_1 - p_2 (E_1 + E_2)  
\end{align}
which is the opposite of the energy of the completely passified state \cite{Allahverdyan2004}, while the energy for unit subsystem obtainable from performing entangling operations is
\begin{align}
  \frac{\langle H \rangle_2}{2}=&\begin{cases}
    E_2p_0^{2}+(E_1+2E_2)p_0p_1+(E_2-E_1)p_1^{2}+&\\2E_1p_0p_2-(E_1+\frac{3}{2}E_2)p_1p_2-(E_1+\frac{5}{2})p_2^{2},  &\\\hspace{4.5cm} \text{if $p_0p_2 \le p_1^{2}$}.\\\newline\\
    E_2p_0^{2}+(E_1+2E_2)p_0p_1+(\frac{E_1}{2}+E_2)p_0p_2+&\\ \frac{E_1}{2}p_1^{2}-(E_1+\frac{3}{2}E_2)p_1p_2-(E_1+\frac{5}{2})p_2^{2}, &\\ \hspace{4.5cm} \text{if $p_0p_2>p_1^{2}$}.
  \end{cases}
\end{align} 

\begin{figure}
    \centering
    \includegraphics[width=\columnwidth]{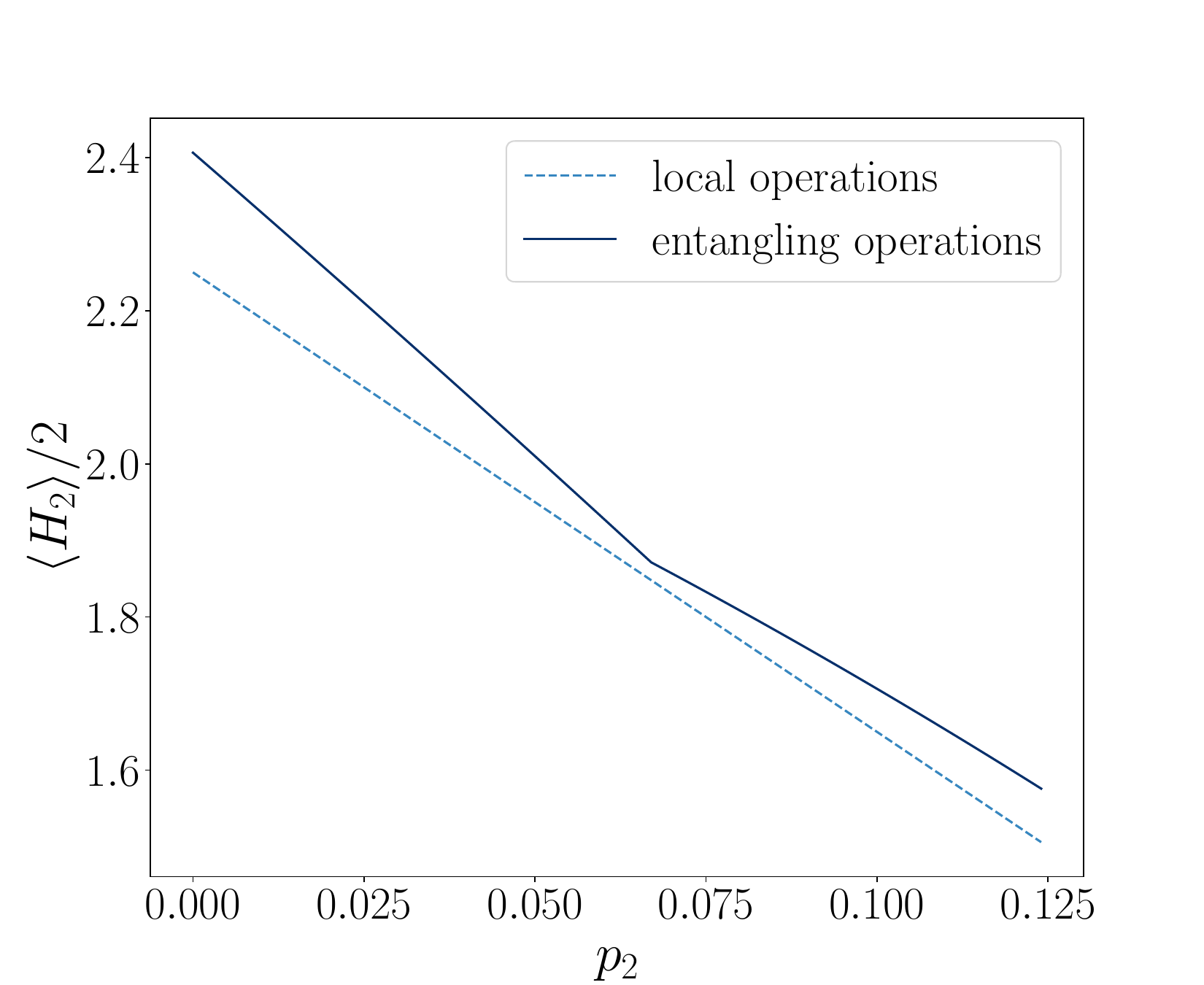}
    \caption{Energy per site for a five-level system when player $A$ can do two-site operations (continuous blue line) and when it can only do single site operations (dashed light blue line). The change in direction of the orange line represents the point from which $p_0p_2> p_1^{2}$. }
    \label{fig:ergotropy}
\end{figure}

Indeed, for two five-level systems, the maximum expectation value of the energy per particle $\avH_2$ obtainable depends on the exact numerical values of $E_1$ and $E_2$ and on whether $p_0 p_2$ is larger or smaller than $p_1^2$. To show a more concrete example, we consider a scenario for which $p_0=0.5$ and we vary $p_2$ between $0$ to $0.12$, while $p_1$ is constrained by the fact that the trace is $1$. We also take $E_1=1$ and $E_2=4$ and we plot, in Fig.~\ref{fig:ergotropy}, the energy per particle versus $p_2$ when you can do two site operations $\avH_2/2$ (continuous blue line) and when you can only do single site operations $\avH_1$ (dashed light blue line). This shows that there are regions in the parameter space in which one can get extra gain from doing entangling operations on more than one system, even though considering just single site measurement operators. 

What we have discussed until now shows that player A has the potential, when acting second, not only to maximize the energy from each $\rho_i$, but to actually extract, from using unitaries on two five-level systems, even more than what she would get from acting only on one after undoing what player B has first implemented on the systems. 
Hence, in principle player A may not even need to be able to apply a perfect defence which gives player B a perfectly mixed state. In short, player A would be able to obtain $\avH_{BA}>\avH_{AB}$ even without doing anything when acting first. 

\begin{figure}
    \centering  \includegraphics[width=\columnwidth]{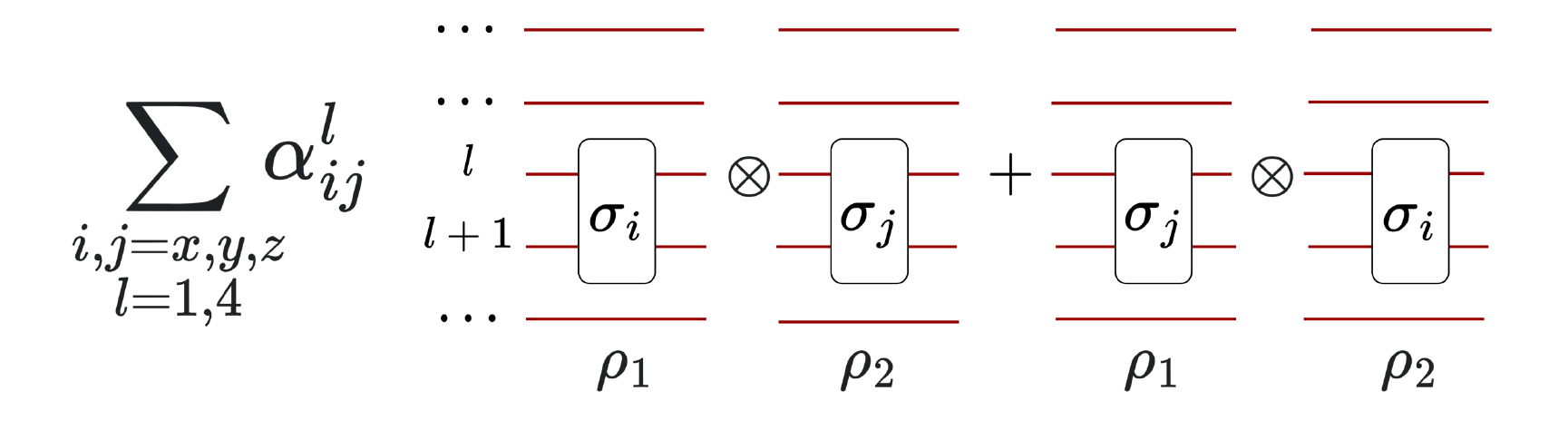}
    \caption{Graphical depiction of the representation of the generators of the unitary transformation used to find the maximum entanglement entropy reachable for the pair of mixed 5-level systems.}
    \label{fig:unitary_5_levels}
\end{figure}

Interestingly, player A can actually also implement a perfect defence for two 5-level systems by preparing the state $\ket{\psi^{+}_l}=\frac{1}{\sqrt{l}}\sum\limits_{i=0}^{l-1}\ket{i}_A\otimes \ket{i}_B$ for $l=5$.  
However, this does not automatically imply that A can achieve a perfect defence from an initally provided mixed state. 
To do this we prepare an algorithm which maximazes the single-site entropy for two five-level systems under the effect of unitaries acting on both of them. The unitaries are parametrized as in Fig,~\ref{fig:unitary_5_levels} by considering a composition of four-level unitaries acting on the 25-level system which results from the two five-level systems.  To ensure the possibility to reach the maximum entropy, we used a symmetrized version of the four-level unitaries such that they act on the same two levels for each of the five-level systems. For example the unitary acts on levels 1 and 2 for both systems. Then on levels 2 and 3, then 3 and 4, followed by 4 and 5 and then back to 3 and 4 all the way back to 1 and 2. Furthermore, each four-level unitary is represented by 9 different parameters and not the total 15 of them. This is because we consider each four-level unitary $U_4$ with generators $G_{ab}$ composed with symmetric operators on the two system 
$U=\prod_l\exp(-\im \sum_{i,j}\alpha^l_{ij} G_{ij})$ 
where $\alpha^l_{ij}$ are real numbers which parametrize the unitary. More precisely, we consider the generators $G_{ii} = \sigma_i\otimes\sigma_i$ for $i=x,y,z$ and $G_{ij} = \sigma_i\otimes\sigma_j + \sigma_i\otimes\sigma_j $ for $i\ne j$ and both $i,j = x,y,z$ or $0$ for the identity matrix. Note that the single generators $\sigma_i$ act on two levels of each system only, levels $l$ and $l+1$. In doing so, we iteratively optimize the coefficients of the unitaries to maximize the single site entropy, and we find that the Von Neumann entropy converges to the value expected from the infinite temperature state $\log_5(5)=1$. 
It is thus possible for player A to operate a perfect defence when playing with two five-level systems, and when player A acts as the second player, she can potentially get more energy than trivially expected if acting on each single system separately. 
  
\section{Conclusions}\label{sec:conclusions} 
We have shown that in a competitive, turn-based, quantum game between two players who can only apply unitaries on the same many-body quantum register, the player who can operate on more qubits can have a significant advantage. The advantage is maximized in the case of AME states when one player can entangle at least twice the number of qubits than the other. Furthermore, the advantage can be more important when dealing with systems with more than two levels as they can be turned into AME states for a larger number of components. When dealing with mixed states, for systems with more than two levels, it is also possible to obtain extra gains from the player with a quantum advantage, in a similar way as discussed in the field of quantum batteries \cite{AlickiFannes2013}. 
For cases in which the player with quantum advantage cannot reach an AME, e.g. because there are no AMEs for that system size
, one defensive strategy for the player with quantum advantage is to use random unitaries which asymptotically lead the system to approach an AME. However, it will still be very important to have a significant quantum advantage, i.e. being able to operate coherently on more sites than the other player. 

Possible future work could consider the case in which the two players submit a certain generator of unitary dynamics to a referee who adds them up and uses the result to generate the unitary evolution. This is a fundamentally different framework, and even if the set of generators used by the two players is very different, e.g. different size of support, the game can end, on average, in a tie. 
Other future works can study the case in which players A and B, while acting on the same register, can only pick between different, and limited, sets of unitaries.
Another possibility is to study the case in which players can also operate on the register with dissipative channels. Closer inspection on similarities and differences between classical and quantum setups could be studied, both considering only doubly stochastic matrices and dissipative scenarios. In this case the differences between classical and quantum scenario can be significantly reduced.

{\it Acknowledgments:} We are grateful to G. Piliouras for fruitful discussions. D.P. would also like to thank R. Jain, J.I. Latorre and H.-K. Ng for fruitful discussions. A.V would like to thank W. Lin for useful discussions. D.P. acknowledges support from joint Israel-Singapore NRF-ISF Research grant NRF2020-NRF-ISF004-3528. A.V. acknowledges support from the Quantum Engineering Programme NRF2021-QEP2-02-P05. 

\bibliographystyle{apsrev4-1}
\bibliography{references}

\appendix 

\section{Example for $N=3$ and $N_B=2$}\label{app:example} 
  
Let us consider the  example of $N=3$ and $N_B=2$. We write  
$$|\psi_B\ra=U_{12}\otimes U_3|\psi_A\ra$$ 
where $\ket{\psi_{A/B}}$ is the wavefunction of the system after player A or B acts on it. 
After player B applies his unitaries, the expected value  of $H$ is 
\begin{align}
\la H\ra_{AB} =&\tr\left(  I_1\otimes Z_2+ Z_1\otimes I_2, \tr_3(\ket{\psi_B}\bra{\psi_B})\right) \nonumber \\
&+\tr\left( Z, tr_{12}(\ket{\psi_B}\bra{\psi_B})\right)  \nonumber  
\end{align}
Note that 
 $$ \tr_3(\ket{\psi_B}\bra{\psi_B})=U_{12} \tr_3\left(\ket{\psi_A}\bra{\psi_A}\right)U_{12}^\dagger$$ 
 and 
$$ \tr_{12}(\ket{\psi_B}\bra{\psi_B})=U_{3} \tr_{12}\left(\ket{\psi_A}\bra{\psi_A}\right)U_{3}^\dagger$$
where 
\begin{align}
\tr_3(\ket{\psi_A}\bra{\psi_A})=\begin{bmatrix}
         \lambda_1 & 0 & 0 &0\\ 
         0& \lambda_2  & 0 &0\\
         0& 0 &0  &0 \\
         0& 0 & 0 &0\\
     \end{bmatrix} 
\end{align} 
and 
\begin{align}
\tr_{12}(\ket{\psi_A}\bra{\psi_A})=\begin{bmatrix}
         \lambda_1 & 0 \\ 
         0& \lambda_2  \\
     \end{bmatrix} 
\end{align} 

Without loss of generality, we consider $1\ge \lambda_1 \ge\lambda_2\ge 0 $.   

Player A needs to figure out what is the worst-case scenario for her, or equivalently,  what is the best  case scenario for player B. From  player B's perspective, he needs to choose the unitaries $U_{12}, U_3$ that minimize the energy, i.e.,  he needs to solve  the following optimization problem
\begin{align}
    \min_{U_{12}, U_3} & \left[\tr\left(  (I_1\otimes Z_2+ Z_1\otimes I_2) U_{12} \tr_3(\ket{\psi_A}\bra{\psi_A})U_{12}^\dagger\right) \right.\nonumber \\ 
    &\left.+\tr\left( Z_3 U_{3} \tr_{12}(\ket{\psi_A}\bra{\psi_A})U_{3}^\dagger\right) \right] \nonumber 
\end{align}
Note that this optimization is  separable with respect to $U_{12}, U_3$, so player B  needs to solve two separate problems for $U_{12}$ and $U_3$ respectively. 
To do so, player B will make the states passive for both the two-qubit system, and the one-qubit. 
For these two cases, the energies are $-2$, $0$ (degenerate) and $2$, while for the one qubit the energies are $-1$ and $1$. 
Hence the lowest energy obtainable for player B is $-2\lambda_1$ for the two-qubit portion, and $-\lambda_1+\lambda_2$ for the one-qubit portion. 
This gives 
\begin{align}
    \la H\ra_{AB} & = -3\lambda_1 + \lambda_2 \nonumber \\ 
    & = -4 \lambda_1 +1, 
\end{align}
where we have used that $\lambda_1+\lambda_2=1.$ 
It is thus clear that the best strategy for player A is to try to set $\lambda_1$ as close as possible to $1/2$, which results in $\la H\ra_{AB}=-1$.

\end{document}